\documentclass[conference]{IEEEtran}
\IEEEoverridecommandlockouts
\usepackage{cite}
\usepackage{amsmath,amssymb,amsfonts}
\usepackage{algorithmic}
\usepackage{graphicx}
\usepackage{textcomp}
\usepackage{xcolor}
\def\BibTeX{{\rm B\kern-.05em{\sc i\kern-.025em b}\kern-.08em
    T\kern-.1667em\lower.7ex\hbox{E}\kern-.125emX}}

\usepackage{listings}
\usepackage{color}

\definecolor{codegreen}{rgb}{0,0.6,0}
\definecolor{codegray}{rgb}{0.5,0.5,0.5}
\definecolor{codepurple}{rgb}{0.58,0,0.82}
\definecolor{backcolour}{rgb}{0.95,0.95,0.92}

\lstdefinestyle{mystyle}{
	backgroundcolor=\color{backcolour},   
	commentstyle=\color{codegreen},
	keywordstyle=\color{magenta},
	numberstyle=\tiny\color{codegray},
	stringstyle=\color{codepurple},
	basicstyle=\footnotesize,
	breakatwhitespace=false,         
	breaklines=true,                 
	captionpos=b,                    
	keepspaces=true,                 
	numbers=left,                    
	numbersep=5pt,                  
	showspaces=false,                
	showstringspaces=false,
	showtabs=false,                  
	tabsize=2
}

\begin{document}

\title{Interference analysis of shared last-level cache on embedded GP-GPUs with multiple CUDA streams
}

\author{\IEEEauthorblockN{Gianluca Brilli, Paolo Burgio}
\IEEEauthorblockA{
\textit{University of Modena and Reggio Emilia, Italy}\\
\{gianluca.brilli, paolo.burgio\}@unimore.it}
}

\maketitle

\begin{abstract}
In modern heterogeneous architectures, the access to data that the application needs is a key factor, in order to make the compute task efficient, in terms of power dissipation and execution time. The new generation SoCs are equipped with large LLCs, in order to make data access as efficient as possible. However, these systems introduce a new level of complexity in terms of the system's predictability, because concurrent tasks must compete for the same resource and contribute to generating interference between them. This paper aims to provide a preliminary qualitative analysis in terms of interference degree that is generated when several concurrent streams are in execution, for example one that performs useful computing tasks and one that generates interference. Specifically, we tested two important primitives: \textit{vadd} and \textit{gemm}, respectively subjected to interference with: \textit{i)} a concurrent kernel that performs read from shared memory. \textit{ii)} concurrent stream that performs host-to-device memory copy.
\end{abstract}


\section{Introduction, and motivation for this work}
\begin{figure}[h!]
	\includegraphics[width=\columnwidth]{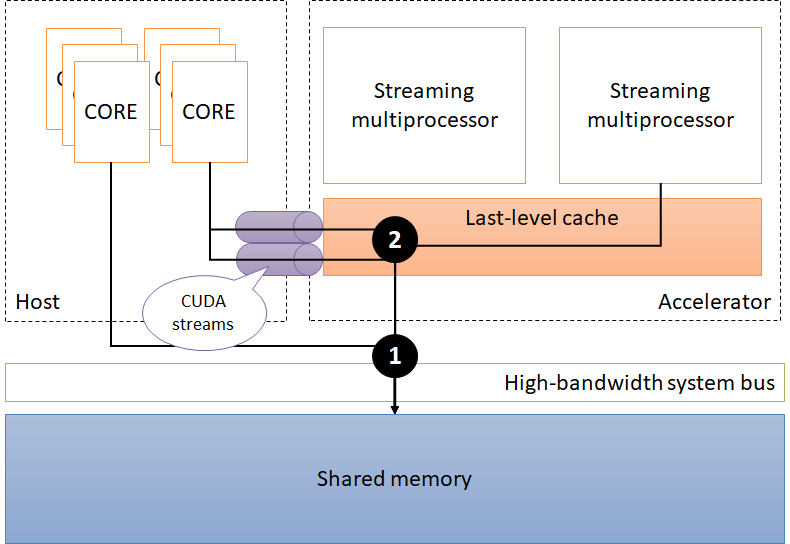}
	\centering
	\caption{Reference iGPU architecture with key architectural bottlenecks}
	\label{fig:tx2_arch}
\end{figure}

The increasing demand for high-performance computational capabilities at low size-weight and power (SWaP) of modern embedded systems paved the way to the adoption of heterogeneous computing platforms with multi-core host and many-core accelerators.
Especially, integrated GPGPUs (iGPUs)\cite{xavier, tegra_x2} are today's preferred to other acceleration paradigms, e.g., based on FPGAs or application-specific integrated circuits (ASICs), in applications with data-parallel workloads, such as computer vision and AI systems employing deep neural networs.
This is the case of advanced automotive systems\footnote{Figure~\ref{fig:tx2_arch} shows a simplified block diagram of a NVIDIA TX2, where an esa-core host shares memory banks with two CUDA streaming multiprocessors (SM) of the Pascal family.}, where AI/DNN are increasingly being adopted as reference for building partly- or fully- automated vehicles of tomorrow.
Unfortunately, these systems demand not only for high peek performance, but also --and especially-- worst case performance, and the increased architectural complexity of modern iGPUs makes it extremely cumbersome to perform an effective non-pessimistic worst-case timing analysis of system.
Recently, researchers~\cite{cavicchioli_ETFA17, mancuso_ECRTS17, yun_TC7} proved that the main source of unpredictability in such systems are contentions on shared resources, such as memory banks, but yet only few works~\cite{bjorn_DATE19} focused on shared GPU last-level cache (LLC) which also is a major source of contention, \textit{that affects both host and accelerator complexes}.
The reason for this lack of material is that, hardware providers (in this case, NVIDIA) are too often reluctant to disclose the internals of their highly-optimized architectures and memory drivers, forcing researcher to a huge effort of reverse engineering for understanding them~\cite{bjorn_DATE19}.
This is also interesting because \textit{last-level-caches are the closest shared resources between cores}, hence they are affected by the whole memory traffic due to local caches misses, and deserve a special attention.

\textbf{CUDA streams.}
One of the main performance booster, when adopting a host-accelerator paradigm, is the possibility of overlapping multiple computation kernels and data transfers between the CPU and the GPU.
In NVIDIA GPGPUs, this is possible thanks to the abstraction of \textit{CUDA streams}, where both execution and data transfer request are issued from the application control running on the host.
Unfortunately for RT engineers, CUDA streams introduce an additional level of parallelism, further increasing system complexity, and we will show how the complex mechanism for stream management implemented in platform drivers \textit{enables an additional source of contention in the system}, negatively affecting predictability, because they create interference not only on the shared memory, but also on last-level cache.
Circles with numbers in Figure~\ref{fig:tx2_arch} highlight the two main contention points (LLC and memory) in the considered system.

\textbf{Platform modeling in industry.}
Another issue stems from the fact that industrial-grade frameworks for software development, such as Amalthea~\cite{amalthea} for the automotive domain, too often rely on simplified platform model, practically inapplicable and ineffective with the complex structure of iGPUs.
Indeed, there is no standard approach to modeling both the implicit memory contention between host cores and GPU cores.
Of course, the situation gets even worse when CUDA streams are included in the picture. 
Indeed, this year's WATERS challenge only focuses on single-stream applications.

Our work wants to be the first one in analyzing and modeling, not only analytically but also with empirical evidence, the contention on LLC introduced by the adoption of multiple CUDA streams.

The rest of the paper is structured as follows: in section II, we provide some ideas regarding the implementation of the interference benchmarks and some expected results. Section III shows the graphs obtained by running the experiments described in Section II on top a representative embedded iGPU, the NVIDIA Tegra X2 and details the results. Finally in Section IV we summarize the experiments carried out in this research and we conclude the paper. 
\section{Implementation Details}

In this section we provide some details related to the benchmark suite used in the experiments.
In particular we decided to measure the slowdown due to the interference on LLC on two basic operations of Linear Algebra, respectively the sum of vectors and the multiplication of matrices ( \textit{vadd} and \textit{gemm} ).
From the point of view of the experiments carried out, we quantified the level of cache interference by measuring the performance of the aforementioned kernels with and without interference, in terms of execution time, because unfortunately the metric \textit{l2\_l1\_read\_hit\_rate} is not implemented on our target platform. In this way we compared the execution times of the two compute-kernels with and without interference. The slowdown obtained, is obviously due to the LLC replacement and the consequent high cost of access to the \textit{Shared Memory} outside the chip.

In terms of benchmarks, we decided to test the following two cases:

\textit{i) Compute Kernel and Interference Kernel}: in this first test case we performed a comparison between a compute kernel running on a single \textit{Streaming Multiprocessor (SM)} in isolation (ie. without interference), compared to the same kernel in a concurrent execution with a kernel that reads data from memory, running on a parallel stream, mapped to the second SM inside the Tegra X2 SoC.

\begin{lstlisting}[caption={Interference kernel implementation.},captionpos=b\label{lst:a_label},language=C]
while (runs --) {
	idx = threadIdx.x * stride;
	while (idx < n) {
		w[idx] = r[idx];
		idx   += blockDim.x;
	}
}
\end{lstlisting}

In listing 1 we can see the simple idea behind  the interference kernel used in this work. In particular we can notice the \textit{stride} parameter, which determines the access pattern of the threads to the LLC.
\begin{figure}[h!]
	\includegraphics[width=\columnwidth]{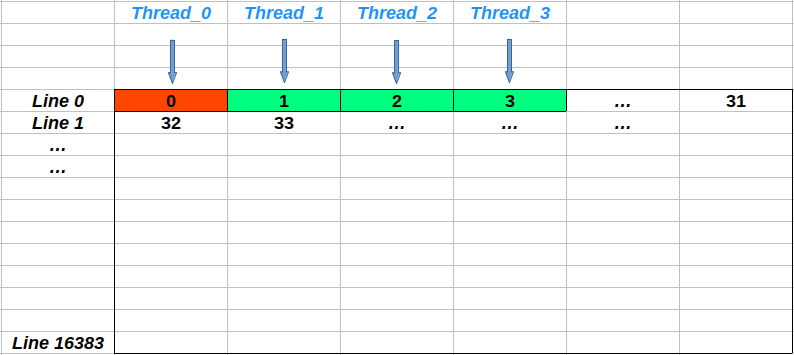}
	\centering
	\caption{Thread access pattern to LLC. The case reported in this figure is a \textit{coalesced} access, that we have when the \textit{stride} is equal to $1$. In this case only the first thread causes a cache miss (the red cell). }
	\label{fig:label_f}
\end{figure}

In figure \ref{fig:label_f} we can see what happens inside the GPU LLC from a graphical point of view, varying the stride.
In particular if it is less than the cache line size ($ < 32 $) we expect the generated interference is "relatively low", because some threads will perform cache hit on the LLC. Similarly, by increasing this parameter beyond the LLC line size ($ > 32 $), we experimented a similar behavior, in which some cache lines will not be affected by the interference memory accesses.
Finally in the case of strided accesses of the same size as the cache line ($ = 32 $), we have the maximum possible interference that we can generate involving a single SM \footnote{The maximum number of threads that can be launched in a block is hardware-limited to $1024$ threads, summing each dimension of the grid.}, then in this last case, each thread could potentially perform a cache miss, triggering a load from the main memory.

\textit{ii) Compute Kernel and Copy Engine}: in this last comparison we focused on the behavior generated by concurrently launch a compute-kernel, running on top of one of the two SMs, together with a concurrent CUDA-stream that takes care of performing memory copy \textit{host-to-device}, through the \textit{Copy Engine}. This type of copy, moves the target memory, which is stored in an area addressable by the Host complex, inside a region that is visible by the GPU, and also prefetches the data needed on the GPU LLC, in order to bring this data closer to the \textit{Compute Engines}. Within our benchmark, these memory copies involve all or some parts of cache lines, in particular the cache regions are kept under interference throughout the computation time of the concurrent kernel.

\begin{lstlisting}[caption={Concurrent memory copy stream.},captionpos=b\label{lst:a_label},language=C]
while(runs --) {
	cudaMemcpyAsync(w_gpu, r_gpu, LINE_SIZE * cache_lines, cudaMemcpyHostToDevice, s1);
	cudaStreamSynchronize(s1);
}
\end{lstlisting}

Listing 2 shows in detail the benchmark idea described above. In particular we used the \textit{CudaMemcpyAsync}, for executing memory copy on top a concurrent stream, different from the default one. Through the \textit{cache\_line} parameter we can control the number of cache lines under interference, ranging from one single line up to the whole cache size.
\section{Experimental Results}

In this section we report the experimental results from two kernels in isolation, compared to a concurrent execution with the two aforementioned interference cases. The target computing platform is the NVIDIA Tegra X2 \cite{tegra_x2}, a \textit{System-on-Chip (SoC)} that embeds an esa-core processor, in a Big.SUPER configuration, composed of four ARMv8 Cortex A57 and two proprietary NVIDIA Denver. The GPU is composed of two Streaming Multiprocessors (SMs) belonging to the \textit{Pascal} family, summing up $256$ CUDA Cores @ $854-1465$MHz in total. As mentioned before, we used one SM to perform useful computation and the other one for doing memory interference.
Regarding the memory hierarchy, we have $8$ GB of LPDDR4, shared between the CPU cluster and the GPU, with a bus width of $128$ bits and a bandwidth of $58.4$ GB/s. Regarding the cache system we have a L2 cache, shared between the host complex and a L2 cache for the \textit{Pascal} GPU, the second one is composed of $16384$ cache lines of size $32$ bytes each, summing up $512$KB in total.
\begin{figure*}[h!]
	\includegraphics[width=\textwidth]{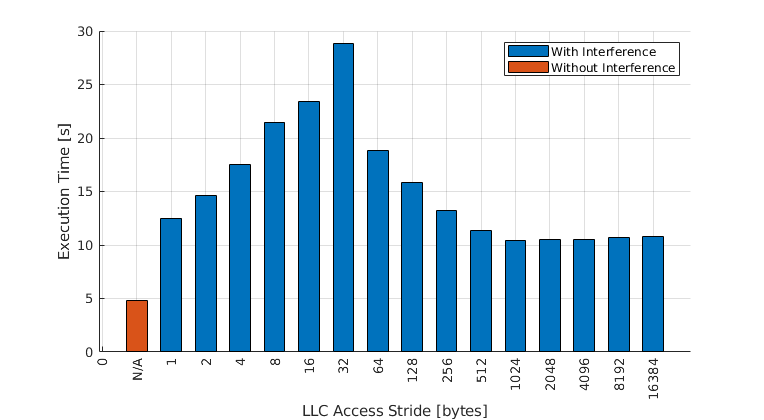}
	\caption{Comparison in terms of execution time of the \textit{vadd} kernel in isolation (red bar), compared to the same kernel running concurrently with an interference kernel (blue bars).}
	\centering
	\label{fig:vadd_kernel}
\end{figure*}
\begin{figure*}[h!]
	\includegraphics[width=\textwidth]{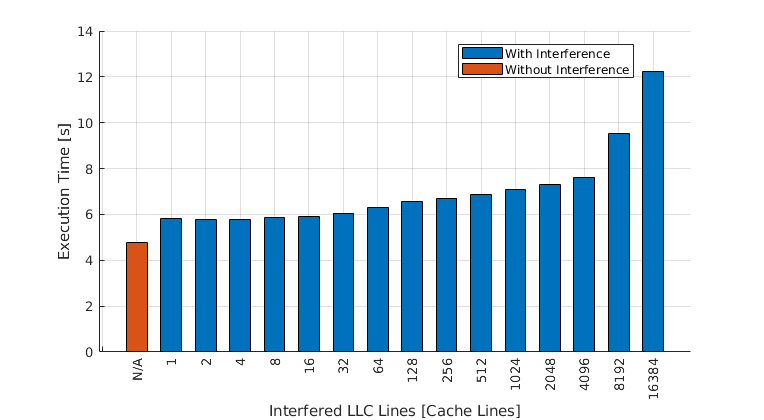}
	\caption{Comparison in terms of execution time, between the \textit{vadd} kernel in isolation (the red bar) and the same kernel in execution with a concurrent stream that performs memory copy \textit{host-to-device} (the blue bars).}
	\centering
	\label{fig:vadd_memcpy}
\end{figure*}
\begin{figure*}[h!]
	\includegraphics[width=\textwidth]{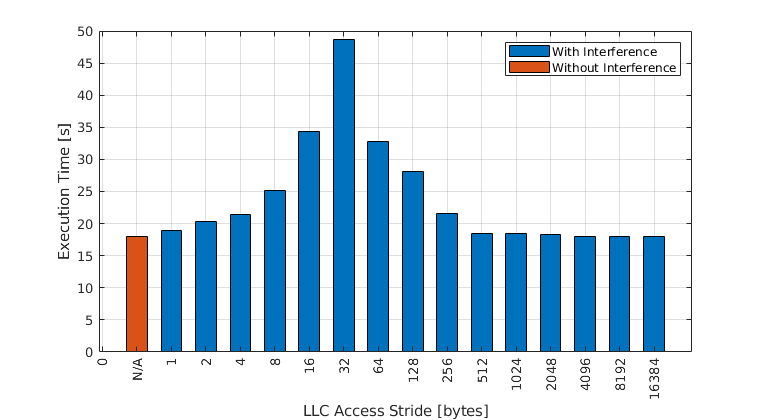}
	\caption{Comparison in terms of execution time of the \textit{gemm} kernel in isolation (red bar), compared to the same kernel running concurrently with an interference kernel (blue bars).}
	\centering
	\label{fig:gemm_kernel}
\end{figure*}
\begin{figure*}[h!]
	\includegraphics[width=\textwidth]{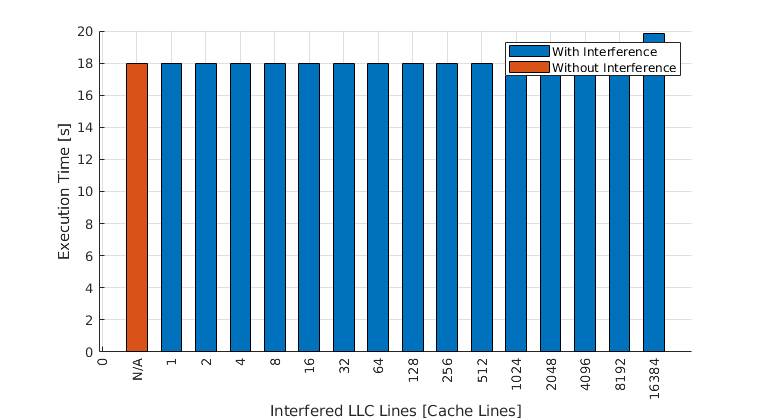}
	\caption{Comparison in terms of execution time, between the \textit{gemm} kernel in isolation (the red bar) and the same kernel in execution with a concurrent stream that performs memory copy \textit{host-to-device} (the blue bars).}
	\centering
	\label{fig:gemm_memcpy}
\end{figure*}

In figure \ref{fig:vadd_kernel} we can see a comparison of the execution time needed by the \textit{vadd} kernel, respectively with and without interference.
In particular from this first benchmark we experimented an execution time of about $5$ seconds for the baseline version (without interference). Subsequently by launching the interference kernel and varying the \textit{stride} parameter we can notice a linear growth in terms of execution time, until reaching the dimension of the LLC line-size. In this case we have reached the maximum number of possible cache misses and the consequent loads from the main memory, which we can generate with a single SM, composed of one CUDA block of $1024$ concurrent threads. The increase in execution time experimented is proportional to a $6\times$ factor from the baseline version. Further increasing the \textit{stride} parameter even beyond the LLC line-size we have a reduction of the interference factor, up to about $2\times$ compared to the baseline version. Specifically, we observed that the execution time seems to converge towards a specific value. The decrease in execution time is due to the fact that increasing the \textit{stride} parameter beyond the line size of the LLC, we are indeed skipping some cache lines and consequently generating less interference.

In figure \ref{fig:vadd_memcpy} we can see a comparison of the same compute kernel, with a concurrent execution of a CUDA stream that performs memory copy \textit{host-to-device}, exploiting the \textit{Copy Engine}. The comparison is always measured in terms of execution time and in this case it is carried out by gradually increasing the number of cache lines affected by the memory copy stream, ranging from a single cache line, up to $16384$ cache lines (the full size of the LLC). In this case we have an increase of about a $1.2\times$ factor in the case of a single cache line interfered, up to a factor of $2.4\times$ when we have interference on the whole cache.

The same test modalities are shown in figure \ref{fig:gemm_kernel} and \ref{fig:gemm_memcpy}, in particular in figure \ref{fig:gemm_kernel} we can see the growth of terms of execution time due to the interference generated by the concurrent kernel in execution on the second SM. Also in this case we can notice a linear growth of the execution time, by increasing the \textit{stride} parameter, until we reach the line-size of the LLC ( $32$-bytes ). In case of the \textit{gemm} task we have an increase of about a $3\times$ factor of the execution time respect to the baseline reference. Beyond this point, also in this case, follows a decrease in the total execution time, until reaching a fixed value, similar to the baseline case.

Finally in the last proposed comparison, shown in figure \ref{fig:gemm_memcpy}, we note that the effect of interference due to memory copy \textit{host-to-device} does not have a particularly interference impact on the total execution time, unlike what happens in the case of concurrent kernels.
\section{Conclusion}
In this paper we presented, through a sequence of benchmarks, some interference effects on \textit{Last-Level Cache (LLC)}, due to the execution of an interference kernel, mapped on a concurrent \textit{Streaming Multiprocessor (SM)} and by means of memory copy \textit{host-to-device}. The tests were carried out on a device that nowadays can be considered as the state of the art regarding high performance iGPU, the NVIDIA Tegra X2 SoC. Through the analysis of the results obtained we were able to infer the size of the single cache line, a fundamental parameter to be known if we intend to develop embedded software that makes efficient use of caches. We also highlighted that interference on LLC, in case of kernels running on separate SMs, has a very high impact on the number of accesses to the main memory, which translates into \textit{i)} a considerable increase in total execution time, equal to a factor of about $6\times$ in the case of the \textit{vadd} and about $3\times$ for the \textit{gemm}. \textit{ii)} difficulty in making predictable the \textit{worst-case execution time (WCET)}. This justifies and supports the need for adopting predictable models, in which several tasks must synchronize the memory access phases.

\bibliographystyle{abbrv}
\bibliography{biblio}
	
\end{document}